\documentclass[showpacs,showkeys,aps,prb,onecolumn]{revtex4}
\usepackage[english]{babel}
\usepackage{epsfig}
\usepackage{graphicx}
\usepackage{amsmath}
\usepackage{amssymb}
\usepackage{subfigure}
\usepackage{bm}
\usepackage{subfigure}

\begin{document}

\title{Ground state energy of $N$ Frenkel excitons}
\author{Walter Pogosov (1,2) and Monique Combescot (1)}
\affiliation{(1) Institut des NanoSciences de Paris, Universite Pierre et Marie Curie,
CNRS, Campus Boucicaut, 140 rue de Lourmel, 75015 Paris}
\affiliation{(2) Institute for Theoretical and Applied Electrodynamics, Russian Academy
of Sciences, Izhorskaya 13/19, 125412 Moscow}
\date{\today }

\begin{abstract}
By using the composite many-body theory for Frenkel excitons we have
recently developed, we here derive the ground state energy of $N$ Frenkel
excitons in the Born approximation through the Hamiltonian mean value in a
state made of $N$ identical $\mathbf{Q=0}$ excitons. While this quantity
reads as a density expansion in the case of Wannier excitons, due to
many-body effects induced by fermion exchanges between $N$ composite
particles, we show that the Hamiltonian mean value for $N$ Frenkel excitons
only contains a first order term in density, just as for elementary bosons.
Such a simple result comes from a subtle balance, difficult to guess a
priori, between fermion exchanges for two or more Frenkel excitons appearing
in Coulomb term and the ones appearing in the $N$ exciton normalization
factor - the cancellation being exact within terms in $1/N_{s}$ where $N_{s}$
is the number of atomic sites in the sample. This result could make us
naively believe that, due to the tight binding approximation on which
Frenkel excitons are based, these excitons are just bare elementary bosons
while their composite nature definitely appears at various stages in the
precise calculation of the Hamiltonian mean value.
\end{abstract}

\pacs{71.35.-y, 71.35.Aa}
\maketitle

\section{Introduction}

Quite recently, we have constructed a many-body theory for Frenkel excitons 
\cite{paper1,paper2} similar to the one we have developed for Wannier
excitons. In these theories, the composite nature of the particles is
treated exactly through a set of "Pauli scatterings" for carrier exchanges
between two excitons in the absence of carrier interaction. Exchanges
between two or more excitons, which control the many-body physics of these
composite particles, are visualized and readily calculated by using a new
diagrammatic representation, called "Shiva diagrams" in relation with their
multiarm structure.

Alternative approaches which allow one to take into account the Pauli
exclusion principle in the case of Frenkel excitons are based on expressing
the fermionic Hamiltonian as an infinite series expansion in creation and
destruction operators for excitations on atomic sites. In the formalism
proposed by Agranovich and coworkers \cite{AgrGal,Agranovich,AgrTos}, these
operators obey bosonic commutation rules, while in the method developed by
Mukamel and coworkers \cite{CherMuk,Mukamel,Muk1,Muk2,Mukamel-review}, the
commutation rules for these operators are not bosonic. The precise forms of
the resulting Hamiltonian expansions produced by these two approaches also
differ from each other. In contrast to these approaches, the composite boson
many-body theory we have constructed \cite{paper1,paper2} avoids the
difficulty of deriving Hamiltonian infinite series expansions with an
increasing number of operators \cite{Monique-seria}, the processes involving
any number of excitons being in our formalism treated on the same footing.
This enables us, for the first time in our best knowledge, to
self-consistently study interactions between an arbitrary number of Frenkel
excitons while including the tricky consequence of the Pauli exclusion
principle in an exact way.

The composite nature of the Wannier excitons turns out to be crucial in
their many-body physics. This becomes quite reasonable once we note that the
extension of the electron-hole relative motion in an exciton - which is
expected to control carrier exchanges - is as large as the scale for Coulomb
interaction between excitons, this scale being the exciton Bohr radius $%
a_{x} $ calculated with the semiconductor dielectric constant and the
conduction and valence band effective masses. In contrast, since Frenkel
excitons are made of electrons and holes on the same atomic site, we could
think that, as the extension of their electron-hole relative motion is
essentially zero, fermion exchanges between Frenkel excitons should reduce
to zero; so that these excitons should behave as elementary bosons. In our
previous work on the Frenkel exciton many-body theory \cite{paper2}, we have
shown that, indeed, the closure relation for $N$ Frenkel excitons has the
same $1/N!$ prefactor as the one of $N$ elementary bosons, while for Wannier
excitons, this prefactor is $\left( 1/N!\right) ^{2}$, which is the
signature of the fact that Wannier excitons are made of two independent
fermions. However, for many other basic properties, Frenkel excitons are
composite objects definitely. In particular, they do have Pauli scatterings
for fermion exchange in the absence of fermion interactions. These
scatterings lead to a normalization factor for $N$ identical excitons
different from its $N!$ value for $N$ elementary bosons, by a factor $F_{N}$
which decreases exponentially with $N$. This normalization factor thus
exibits the same "moth eaten" effect induced by Pauli exclusion as the one
found for Wannier excitons.

Fermion exchanges between $N$ Frenkel excitons are found to be controlled by
the dimensionless parameter%
\begin{equation}
\eta =N/N_{s}  \tag{1.1}
\end{equation}%
where $N_{s}$ is the number of atomic sites in the sample. It is of interest
to note that this parameter is actually similar to the dimensionless
parameter $\eta =N\left( a_{x}/L\right) ^{D}$ which controls fermion
exchanges between $N$ Wannier excitons. Indeed, $\left( L/a_{x}\right) ^{D}$%
, like $N_{s}$, is the maximum number of excitons a sample of volume $L^{D}$
can accommodate. Let us stress that, while the expression of $\eta $\ found
for Wannier excitons could make us think that the parameter which controls
fermion exchanges is linked to the extension $a_{x}$ of the electron-hole
relative motion wave function, the expression of $\eta $ for Frenkel
excitons rules out this physical understanding - otherwise $\eta $ for these
excitons would reduce to zero due to the tight-binding approximation
underlying the Frenkel exciton concept.

In the present work on Frenkel excitons, we are interested in the density
dependence of the $N$ ground state exciton energy in the Born approximation.
We reach this quantity through a procedure similar to the one we have used
for Wannier excitons, namely we calculate the Hamiltonian mean value in a $N$
ground state exciton state with momentum $\mathbf{Q}=0$%
\begin{equation}
\left\langle H\right\rangle _{N}=\frac{\left\langle v\right\vert B_{\mathbf{0%
}}^{N}HB_{\mathbf{0}}^{\dag N}\left\vert v\right\rangle }{\left\langle
v\right\vert B_{\mathbf{0}}^{N}B_{\mathbf{0}}^{\dag N}\left\vert
v\right\rangle }.  \tag{1.2}
\end{equation}%
where $B_{\mathbf{Q}}^{\dag }$ creates a Frenkel exciton with momentum $%
\mathbf{Q}$. Let us recall that, while the interacting part of composite
quantum particles cannot be written as a potential so that a "zero order
state in the exciton interaction" cannot be properly defined, the state $B_{%
\mathbf{0}}^{\dag N}\left\vert v\right\rangle $ plays this role since $%
(H-NE_{0})B_{\mathbf{0}}^{\dag N}\left\vert v\right\rangle $ cancels if we
drop all Coulomb scatterings between excitons. As $H$ contains Coulomb
interaction at first order by construction, $\left\langle H\right\rangle
_{N} $ thus corresponds to the ground state energy of $N$ Frenkel excitons
at first order in Coulomb interaction, i.e., in the Born approximation.

We are going to show that, while the $N$ exciton normalization factor reads
as%
\begin{equation}
\left\langle v\right\vert B_{\mathbf{0}}^{N}B_{\mathbf{0}}^{\dag
N}\left\vert v\right\rangle =N!F_{N}  \tag{1.3}
\end{equation}%
where $F_{N}$, calculated in our general paper on the many-body theory of
Frenkel excitons \cite{paper2}, has an expansion in $\eta $ which comes from
fermion exchanges and makes it exponentially decrease with $N$, the
Hamiltonian mean value is found to be \textit{exactly equal} to%
\begin{equation}
\left\langle H\right\rangle _{N}=NE_{0}+\frac{N(N-1)}{2(N_{s}-1)}\gamma _{0}
\tag{1.4}
\end{equation}%
$\gamma _{0}$\ comes from the Coulomb scattering of just \textit{two}
excitons due to both direct Coulomb processes and indirect Coulomb
processes, i.e., the so-called "electron-hole exchange" which is the only
process responsible for the excitation transfer in the case of Frenkel
excitons. Eq. (1.4) thus leads to%
\begin{equation}
\left\langle H\right\rangle _{N}=N\left( E_{0}+\eta \Delta _{0}\right)  
\tag{1.5}
\end{equation}%
\begin{equation*}
\Delta _{0}=\gamma _{0}\left[ 1/2+O(1/N)+O(1/N_{s})\right] 
\end{equation*}%
This shows that the Hamiltonian mean value depends on the dimensionless
parameter $\eta $ associated to exciton density through a first order term
only, corrections being in $N^{-1}$ and $N_{s}^{-1}$,\ but not in $%
N/N_{s}=\eta $.

This has to be contrasted with the Hamiltonian mean value for $N$ Wannier
excitons that we have found to expand as \cite{review}%
\begin{equation}
\left\langle H\right\rangle _{N}=NR_{0}\left( -1+\frac{13\pi }{3}\eta -\frac{%
73\pi ^{2}}{20}\eta ^{2}+\frac{3517\pi ^{3}}{210}\eta ^{3}+O(\eta
^{4})\right)  \tag{1.6}
\end{equation}%
in 3D systems for which the exciton ground state energy is $E_{0}=-R_{0}$\
where $R_{0}=\mu e^{4}/2\varepsilon ^{2}=1/2\mu a_{x}^{2}$ is the exciton
Rydberg. We have explained the presence of terms in density higher than $%
\eta $ by saying that, while $\left\langle H\right\rangle _{N}$ contains one 
$2\times 2$ Coulomb process only by construction, interactions between more
than two excitons exist through fermion exchanges induced by the composite
nature of the particles.

If we now return to Frenkel excitons, the reason for the disappearance of
high order terms is far from trivial. While the tight binding approximation
underlying the Frenkel exciton concept makes most Coulomb exchange processes
between excitons reducing to zero, this is not enough to explain the result
since the denominator of $\left\langle H\right\rangle _{N}$ contains a
density expansion through $F_{N}$. The fact that $\left\langle
H\right\rangle _{N}$ does not contain terms in $\eta ^{n}$\ with $n\geq 2$
actually means that similar $\eta ^{n}$ terms exist in the Coulomb part of $%
\left\langle H\right\rangle _{N}$\ in order to cancel the $\eta ^{n}$ terms
coming from $F_{N}$. This cancellation is however not mathematically exact
since it holds at the order $1/N_{s}$, which indeed reduces to zero in the
large sample limit.

It is clear that the proof of such a tricky cancellation must rely on a very
precise procedure in which the composite nature of the excitons is treated
exactly. This is why the present work heavily uses the composite boson
many-body theory for Frenkel excitons we have just constructed \cite%
{paper1,paper2}. The main results of this theory which is very new, are
recalled in Section II for the reader to possibly follow the rest of the
argument. In Section III, we start with a simpler two-body problem, namely
the calculation of the Hamiltonian mean value for $N=2$ Frenkel excitons,
while in Section IV, we turn to arbitrary $N$. This mean value is compared
to a similar quantity obtained with excitons replaced by elementary bosons, $%
\left\langle H\right\rangle _{N}$ then only having a first order term in
density. In Section V, we conclude.

\section{Main results of the composite exciton many-body theory}

\subsection{General results for composite excitons}

Many-body effects with composite excitons can be derived through a set of
four commutators \cite{review}. The ones linked to fermion exchanges are
obtained from%
\begin{equation}
\left[ D_{mi},B_{0}^{\dagger N}\right] =NB_{0}^{\dagger N-1}\sum_{n}\left\{
\lambda \left( _{m\ i}^{n\ 0}\right) +(m\longleftrightarrow n)\right\}
B_{n}^{\dagger }  \tag{2.1}
\end{equation}%
\begin{equation*}
\left[ B_{m},B_{0}^{\dagger N}\right] =NB_{0}^{\dagger N-1}\left( \delta
_{m0}-D_{m0}\right) 
\end{equation*}%
\begin{equation}
-N(N-1)B_{0}^{\dagger N-2}\sum_{n}\lambda \left( _{m\ 0}^{n\ 0}\right)
B_{n}^{\dagger }  \tag{2.2}
\end{equation}%
where definitions of the deviation-from-boson operator $D_{mi}$ and the
Pauli scattering $\lambda \left( _{m\ i}^{n\ j}\right) $ between two
excitons follow from these equations taken for $N=1$.

In the same way, many-body effects linked to fermion interactions are
obtained from \cite{review}%
\begin{equation}
\left[ V_{i}^{\dagger },B_{0}^{\dagger N}\right] =NB_{0}^{\dagger N-1}\sum
\xi \left( _{m\ i}^{n\ 0}\right) B_{m}^{\dagger }B_{n}^{\dagger }  \tag{2.3}
\end{equation}%
\begin{equation*}
\left[ H,B_{0}^{\dagger N}\right] =NB_{0}^{\dagger N-1}\left(
E_{0}B_{0}^{\dagger }+V_{0}^{\dagger }\right) 
\end{equation*}%
\begin{equation}
+\frac{N(N-1)}{2}B_{0}^{\dagger N-2}\sum \xi \left( _{m\ 0}^{n\ 0}\right)
B_{m}^{\dagger }B_{n}^{\dagger }  \tag{2.4}
\end{equation}%
where definitions of the exciton creation potential $V_{i}^{\dagger }$ and
the interaction scattering between two excitons\ $\xi \left( _{m\ i}^{n\
j}\right) $ follow from these equations taken for $N=1$.

\subsection{Frenkel excitons}

In the case of Wannier excitons, the index $i$ stands for ($\mathbf{Q}_{i}$, 
$\nu _{i}$) where $\mathbf{Q}_{i}$\ is the exciton center of mass momentum
and $\nu _{i}$\ a relative motion index. Due to the tight binding
approximation which in some sense freezes electron on top of hole and makes
Frenkel excitons constructed on electron and hole on the same atomic site,
these excitons do not have relative motion index; so that they only are
characterized by a center of mass momentum $\mathbf{Q}$. Their creation
operator $B_{\mathbf{Q}}^{\dagger }$ reads in terms of free pairs $%
B_{n}^{\dagger }=a_{n}^{\dagger }b_{n}^{\dagger }$ on atomic site $n$ as%
\begin{equation}
B_{\mathbf{Q}}^{\dagger }=\frac{1}{\sqrt{N_{s}}}\sum_{n=1}^{N_{s}}e^{-i%
\mathbf{Q}.\mathbf{R}_{n}}B_{n}^{\dagger }  \tag{2.5}
\end{equation}%
where $N_{s}$ is the number of these atomic sites and $\mathbf{R}_{n}$ their
location. In the same way, free pairs read in terms of excitons as%
\begin{equation}
B_{n}^{\dagger }=\frac{1}{\sqrt{N_{s}}}\sum_{n=1}^{N_{s}}e^{i\mathbf{Q}.%
\mathbf{R}_{n}}B_{\mathbf{Q}}^{\dagger }  \tag{2.6}
\end{equation}%
as easy to check from the system periodicity which makes the following
equations valid%
\begin{equation}
\sum_{n}e^{i\mathbf{(Q}^{\prime }-\mathbf{Q).\mathbf{R}}_{n}}=N_{s}\delta _{%
\mathbf{Q}^{\prime }\mathbf{Q}}  \tag{2.7}
\end{equation}%
\begin{equation}
\sum_{\mathbf{Q}}e^{i\mathbf{Q.(\mathbf{R}}_{n^{\prime }}\mathbf{-\mathbf{%
\mathbf{R}}}_{n}\mathbf{)}}=N_{s}\delta _{n^{\prime }n}  \tag{2.8}
\end{equation}

\subsection{Fermion exchange}

The fact that Frenkel excitons are only characterized by their
center-of-mass momentum, leads to results linked to their many-body
exchanges far simpler than the ones for Wannier excitons. In particular, the
Pauli scattering of two Frenkel excitons is structureless: It only contains
the expected momentum conservation between the "in" excitons ($\mathbf{Q}%
_{1} $, $\mathbf{Q}_{2}$) and the "out" excitons ($\mathbf{Q}_{1}^{\prime }$%
, $\mathbf{Q}_{2}^{\prime }$)%
\begin{equation}
\lambda \left( _{\mathbf{Q}_{1}^{\prime }\ \mathbf{Q}_{1}}^{\mathbf{Q}%
_{2}^{\prime }\ \mathbf{Q}_{2}}\right) =\frac{1}{N_{s}}\delta _{\mathbf{Q}%
_{1}^{\prime }+\mathbf{Q}_{2}^{\prime },\text{ }\mathbf{Q}_{1}+\mathbf{Q}%
_{2}}  \tag{2.9}
\end{equation}%
More generally, the Shiva diagram for carrier exchange between $N$ Frenkel
excitons shown in Fig. 1(b) leads to%
\begin{equation}
\frac{1}{N_{s}^{N-1}}\delta _{\mathbf{Q}_{1}^{\prime }+\mathbf{\ldots }+%
\mathbf{Q}_{N}^{\prime },\text{ }\mathbf{Q}_{1}+\mathbf{\ldots }+\mathbf{Q}%
_{N}}  \tag{2.10}
\end{equation}

By using the Pauli scattering given in Eq. (2.9), we find that the
commutators (2.1, 2) for fermion exchanges reduce to%
\begin{equation}
\left[ D_{\mathbf{Q}^{\prime }\mathbf{Q}},B_{0}^{\dagger N}\right] =\frac{2N%
}{N_{s}}B_{0}^{\dagger N-1}B_{\mathbf{Q-Q}^{\prime }}^{\dagger }  \tag{2.11}
\end{equation}%
\begin{equation}
\left[ B_{\mathbf{Q}},B_{0}^{\dagger N}\right] =NB_{0}^{\dagger N-1}\left(
\delta _{\mathbf{Q}0}-D_{\mathbf{Q}0}\right) -\frac{N(N-1)}{N_{s}}%
B_{0}^{\dagger N-2}B_{\mathbf{-Q}}^{\dagger }  \tag{2.12}
\end{equation}%
From them, it becomes easy to show that the factor $F_{N}$, defined in Eq.
(1.3), which makes $N$ Frenkel excitons differing from $N$ elementary
bosons, obeys the recursion relation%
\begin{equation}
F_{N}=\left( 1-\frac{N-1}{N_{s}}\right) F_{N-1}  \tag{2.13}
\end{equation}%
So that this factor is just given by%
\begin{equation}
F_{N}=\frac{N_{s}!}{(N_{s}-N)!N_{s}^{N}}  \tag{2.14}
\end{equation}%
This compact expression of $F_{N}$ as well as the rather simple commutators
for fermion exchanges between $N$ excitons given in Eq. (2.11, 12) are of
great help to get exact results on Frenkel exciton many-body effects as
shown below.

\subsection{Fermion interaction}

If we now turn to the interaction scattering induced by the Coulomb
interaction which exists between electrons in the various atomic levels, we
have been led to split it as%
\begin{equation}
\xi \left( _{\mathbf{Q}_{1}^{\prime }\ \mathbf{Q}_{1}}^{\mathbf{Q}%
_{2}^{\prime }\ \mathbf{Q}_{2}}\right) =\xi _{coul}\left( _{\mathbf{Q}%
_{1}^{\prime }\ \mathbf{Q}_{1}}^{\mathbf{Q}_{2}^{\prime }\ \mathbf{Q}%
_{2}}\right) -\xi _{transf}\left( _{\mathbf{Q}_{1}^{\prime }\ \mathbf{Q}%
_{1}}^{\mathbf{Q}_{2}^{\prime }\ \mathbf{Q}_{2}}\right) +\xi _{neut}\left( _{%
\mathbf{Q}_{1}^{\prime }\ \mathbf{Q}_{1}}^{\mathbf{Q}_{2}^{\prime }\ \mathbf{%
Q}_{2}}\right)  \tag{2.15}
\end{equation}

Actually, the last part $\xi _{neut}$ of the interaction scattering $\xi $
does not play any role in the many-body physics of Frenkel excitons because,
when inserted in equations like Eq. (2.4), it gives%
\begin{equation}
\sum\limits_{\mathbf{Q}_{1}^{\prime }\mathbf{Q}_{2}^{\prime }}\xi
_{neut}\left( _{\mathbf{Q}_{1}^{\prime }\ \mathbf{Q}_{1}}^{\mathbf{Q}%
_{2}^{\prime }\ \mathbf{Q}_{2}}\right) B_{\mathbf{Q}_{1}^{\prime }}^{\dag
}B_{\mathbf{Q}_{2}^{\prime }}^{\dag }=0  \tag{2.16}
\end{equation}%
Indeed, this neutrality scattering, which comes from the energy $\varepsilon
_{0}$\ we must pay to separate electron from hole, is equal to%
\begin{equation}
\xi _{neut}\left( _{\mathbf{Q}_{1}^{\prime }\ \mathbf{Q}_{1}}^{\mathbf{Q}%
_{2}^{\prime }\ \mathbf{Q}_{2}}\right) =2\varepsilon _{0}\lambda \left( _{%
\mathbf{Q}_{1}^{\prime }\ \mathbf{Q}_{1}}^{\mathbf{Q}_{2}^{\prime }\ \mathbf{%
Q}_{2}}\right)  \tag{2.17}
\end{equation}%
while, due to Eqs. (2.5, 8) and the fact that $B_{n}^{\dag 2}=0$, we do have%
\begin{equation}
\sum\limits_{\mathbf{Q}_{1}^{\prime }\mathbf{Q}_{2}^{\prime }}\lambda \left(
_{\mathbf{Q}_{1}^{\prime }\ \mathbf{Q}_{1}}^{\mathbf{Q}_{2}^{\prime }\ 
\mathbf{Q}_{2}}\right) B_{\mathbf{Q}_{1}^{\prime }}^{\dag }B_{\mathbf{Q}%
_{2}^{\prime }}^{\dag }=\frac{1}{N_{s}}\sum\limits_{\mathbf{q}}B_{\mathbf{Q}%
_{1}+\mathbf{q}}^{\dag }B_{\mathbf{Q}_{2}-\mathbf{q}}^{\dag }=0  \tag{2.18}
\end{equation}

The second part $\xi _{transf}$ of the interaction scattering $\xi $ comes
from the \textit{indirect Coulomb }processes between atomic levels, these
processes insuring the excitation transfer over the whole system in the case
of Frenkel excitons. Such a scattering, quite specific to Frenkel excitons,
can actually be seen as a "transfer assisted exchange", the coupling between
the "in" excitons ($\mathbf{Q}_{1}$, $\mathbf{Q}_{2}$) and the "out"
excitons ($\mathbf{Q}_{1}^{\prime }$, $\mathbf{Q}_{2}^{\prime }$) occurring
through the Pauli scattering of these excitons%
\begin{equation}
\xi _{transf}\left( _{\mathbf{Q}_{1}^{\prime }\ \ \mathbf{Q}_{1}}^{\mathbf{Q}%
_{2}^{\prime }\ \ \mathbf{Q}_{2}}\right) =(\mathcal{V}_{\mathbf{Q}%
_{1}^{\prime }}+\mathcal{V}_{\mathbf{Q}_{2}^{\prime }})\lambda \left( _{%
\mathbf{Q}_{1}^{\prime }\ \ \mathbf{Q}_{1}}^{\mathbf{Q}_{2}^{\prime }\ \ 
\mathbf{Q}_{2}}\right)  \tag{2.19}
\end{equation}%
Its energy-like prefactor is related to the $\mathbf{Q}$ dependent part of
the "out" exciton energy, namely%
\begin{equation}
\mathcal{V}_{\mathbf{Q}}=\sum_{\mathbf{R}\neq 0}e^{-i\mathbf{Q.R}}V_{\mathbf{%
R}}\left( _{1\ 0}^{0\ 1}\right)  \tag{2.20}
\end{equation}%
where $V_{\mathbf{R}}\left( _{\nu _{1}^{\prime }\ \nu _{1}}^{\nu
_{2}^{\prime }\ \nu _{2}}\right) $ is the Coulomb potential between atomic
sites at $\mathbf{R}$, one electron going from the atomic state $\nu _{1}$
to $\nu _{1}^{\prime }$ while the other goes from $\nu _{2}$ to $\nu
_{2}^{\prime }$, the relevant states for Frenkel excitons being the ground
state $\nu =0$ and the first excited state $\nu =1$. This Coulomb coupling $%
V_{\mathbf{R}}\left( _{\nu _{1}^{\prime }\ \nu _{1}}^{\nu _{2}^{\prime }\
\nu _{2}}\right) $ reads in terms of atomic wave functions as%
\begin{equation}
V_{\mathbf{R}}\left( _{\nu _{1}^{\prime }\ \nu _{1}}^{\nu _{2}^{\prime }\
\nu _{2}}\right) =\int d\mathbf{r}_{1}d\mathbf{r}_{2}\varphi _{\nu
_{1}^{\prime }}^{\ast }(\mathbf{r}_{1})\varphi _{\nu _{2}^{\prime }}^{\ast }(%
\mathbf{r}_{2})\frac{e^{2}}{\left\vert \mathbf{r}_{1}-\mathbf{r}_{2}+\mathbf{%
R}\right\vert }\varphi _{\nu _{2}}(\mathbf{r}_{2})\varphi _{\nu _{1}}(%
\mathbf{r}_{1})  \tag{2.21}
\end{equation}

Note that this transfer scattering $\xi _{transf}$ which does not depend on
the momentum transfer between the "in" and "out" excitons, but only on the
momenta of "out" excitons, has similarity with the "photon assisted exchange
scattering" we have recently identified between two polaritons \cite{polar}.

The first part $\xi _{coul}$ of the interaction scattering $\xi $ in Eq.
(2.15) comes from \textit{direct Coulomb} processes between electrons,
between holes and between electrons and holes. As for the Coulomb
scatterings of two Wannier excitons, this scattering depends on the momentum
transfer between the "in" excitons ($\mathbf{Q}_{1}$, $\mathbf{Q}_{2}$) and
the "out" excitons ($\mathbf{Q}_{1}^{\prime }$, $\mathbf{Q}_{2}^{\prime }$)
through%
\begin{equation}
\xi _{coul}\left( _{\mathbf{Q}_{1}^{\prime }\ \mathbf{Q}_{1}}^{\mathbf{Q}%
_{2}^{\prime }\ \mathbf{Q}_{2}}\right) =\frac{\mathcal{W}_{\mathbf{Q}%
_{1}^{\prime }-\mathbf{Q}_{1}}}{N_{s}}\delta _{\mathbf{Q}_{1}^{\prime }+%
\mathbf{Q}_{2}^{\prime },\mathbf{Q}_{1}+\mathbf{Q}_{2}}  \tag{2.22}
\end{equation}%
The prefactor $\mathcal{W}_{\mathbf{Q}}$ is similar to $\mathcal{V}_{\mathbf{%
Q}}$ but for all possible direct Coulomb processes%
\begin{equation}
\mathcal{W}_{\mathbf{Q}}=\sum_{\mathbf{R\neq 0}}e^{-i\mathbf{\mathbf{Q}}.%
\mathbf{R}}\left[ V_{\mathbf{R}}\left( _{1\ 1}^{1\ 1}\right) +V_{\mathbf{R}%
}\left( _{0\ 0}^{0\ 0}\right) -V_{\mathbf{R}}\left( _{0\ 0}^{1\ 1}\right)
-V_{\mathbf{R}}\left( _{1\ 1}^{0\ 0}\right) \right]  \tag{2.23}
\end{equation}

If we now construct exchange Coulomb scattering between two excitons along
the standard procedure, namely%
\begin{equation}
\xi ^{in}\left( _{\mathbf{Q}_{1}^{\prime }\ \ \mathbf{Q}_{1}}^{\mathbf{Q}%
_{2}^{\prime }\ \ \mathbf{Q}_{2}}\right) =\sum_{\mathbf{P}_{1}\mathbf{P}%
_{2}}\lambda \left( _{\mathbf{Q}_{1}^{\prime }\ \ \mathbf{P}_{1}}^{\mathbf{Q}%
_{2}^{\prime }\ \ \mathbf{P}_{2}}\right) \xi \left( _{\mathbf{P}_{1}\ \ 
\mathbf{Q}_{1}}^{\mathbf{P}_{2}\ \ \mathbf{Q}_{2}}\right)  \tag{2.24}
\end{equation}%
we find from Eq. (2.9) and the fact that all scatterings conserve momentum,
that%
\begin{equation}
\xi ^{in}\left( _{\mathbf{Q}_{1}^{\prime }\ \ \mathbf{Q}_{1}}^{\mathbf{Q}%
_{2}^{\prime }\ \ \mathbf{Q}_{2}}\right) =\frac{1}{N_{s}}\delta _{\mathbf{Q}%
_{1}^{\prime }+\mathbf{Q}_{2}^{\prime },\mathbf{Q}_{1}+\mathbf{Q}_{2}}\sum_{%
\mathbf{Q}}\xi \left( _{\mathbf{Q}_{1}+\mathbf{Q}\ \ \mathbf{Q}_{1}}^{%
\mathbf{Q}_{2}-\mathbf{Q}\ \ \mathbf{Q}_{2}}\right)  \tag{2.25}
\end{equation}%
It is then easy to see that, due to Eq. (2.8), the sum over $\mathbf{Q}$
reduces to zero for $\xi _{coul}$ as well as for $\xi _{transf}$ since the $%
R=0$ term is missing from the sums they contain. Consequently,%
\begin{equation}
\xi _{transf}^{in}\left( _{\mathbf{Q}_{1}^{\prime }\ \ \mathbf{Q}_{1}}^{%
\mathbf{Q}_{2}^{\prime }\ \ \mathbf{Q}_{2}}\right) =0=\xi _{coul}^{in}\left(
_{\mathbf{Q}_{1}^{\prime }\ \ \mathbf{Q}_{1}}^{\mathbf{Q}_{2}^{\prime }\ \ 
\mathbf{Q}_{2}}\right)  \tag{2.26}
\end{equation}

In the same way, if we construct the exchange scatterings for $\xi _{transf}$
and $\xi _{coul}$\ between more than two excitons by adding a Coulomb line
on the Shiva diagram shown on Fig. 1b, we find from Eq. (2.10) that the
Coulomb exchange scatterings shown in Fig. 2a in the case of four excitons
reads as%
\begin{equation*}
\sum_{\mathbf{P}_{1}\mathbf{P}_{2}}\frac{1}{N_{s}^{3}}\delta _{\mathbf{Q}%
_{1}^{\prime }+\mathbf{Q}_{2}^{\prime }+\mathbf{Q}_{3}^{\prime }+\mathbf{Q}%
_{4}^{\prime },\mathbf{P}_{2}+\mathbf{P}_{3}+\mathbf{Q}_{3}+\mathbf{Q}%
_{4}}\xi \left( _{\mathbf{P}_{1}\ \ \mathbf{Q}_{1}}^{\mathbf{P}_{2}\ \ 
\mathbf{Q}_{2}}\right) 
\end{equation*}%
\begin{equation}
=\frac{1}{N_{s}^{3}}\delta _{\mathbf{Q}_{1}^{\prime }+\mathbf{Q}_{2}^{\prime
}+\mathbf{Q}_{3}^{\prime }+\mathbf{Q}_{4}^{\prime },\mathbf{Q}_{1}+\mathbf{Q}%
_{2}+\mathbf{Q}_{3}+\mathbf{Q}_{4}}\sum_{\mathbf{Q}}\xi \left( _{\mathbf{Q}%
_{1}+\mathbf{Q}\ \ \mathbf{Q}_{1}}^{\mathbf{Q}_{2}-\mathbf{Q}\ \ \mathbf{Q}%
_{2}}\right)   \tag{2.27}
\end{equation}%
So that, this four-body exchange, as well as all high order exchanges,
contains the same sum over $\mathbf{Q}$ which cancels. This shows that all
Coulomb exchange scatterings between Frenkel excitons of the type of Fig. 2a
reduce to zero. A way to understand the cancellation of these scatterings
could be to say that Frenkel excitons are made of electron on top of hole;
So that the "relative motion wave function" of these pairs reduces to zero.
However this understanding is far too na\"{\i}ve since Pauli scatterings for
carrier exchange between two excitons do exist, as seen from the fact that $%
F_{N}$ is not equal to 1 but exponentially small.

Actually, there are two types of exchange Coulomb scatterings: we can either
have the one of Fig. 2a in which all the "in" and "out" excitons are
constructed on different pairs, but we can also have the one of Fig. 2b in
which one exciton stays made with the same pair. This other Coulomb exchange
scattering which, due to momentum conservation, reads as%
\begin{equation*}
\sum_{\mathbf{P}_{2}}\frac{1}{N_{s}^{2}}\delta _{\mathbf{Q}_{2}^{\prime }+%
\mathbf{Q}_{3}^{\prime }+\mathbf{Q}_{4}^{\prime },\mathbf{P}_{2}+\mathbf{Q}%
_{3}+\mathbf{Q}_{4}}\xi \left( _{\mathbf{Q}_{1}^{\prime }\ \ \mathbf{Q}%
_{1}}^{\mathbf{P}_{2}\ \ \mathbf{Q}_{2}}\right) 
\end{equation*}%
\begin{equation}
=\frac{1}{N_{s}^{2}}\delta _{\mathbf{Q}_{1}^{\prime }+\mathbf{Q}_{2}^{\prime
}+\mathbf{Q}_{3}^{\prime }+\mathbf{Q}_{4}^{\prime },\mathbf{Q}_{1}+\mathbf{Q}%
_{2}+\mathbf{Q}_{3}+\mathbf{Q}_{4}}\xi \left( _{\mathbf{Q}_{1}^{\prime }\ \
\ \ \ \ \ \ \ \ \ \ \ \mathbf{Q}_{1}}^{\mathbf{Q}_{1}+\mathbf{Q}_{2}-\mathbf{%
Q}_{1}^{\prime }\ \ \mathbf{Q}_{2}}\right)   \tag{2.28}
\end{equation}%
does not contain the sum over $\mathbf{Q}$ appearing in Eqs. (2.25) or
(2.27); so that it does not reduce to zero.

Before going further, it can be of interest to note that Coulomb exchange
processes like the one of Fig. 2c a priori exist for Wannier excitons.
However, due to the symmetry between electron and hole in the case of
Wannier excitons, the $2\times 2$ direct Coulomb scattering of these
excitons is such that 
\begin{equation}
\xi \left( _{i\ \ i}^{n\ \ j}\right) =0  \tag{2.29}
\end{equation}%
So that Coulomb exchange scattering like the ones of Fig. 2c reduces to zero
for $m=i$. In contrast, since the wave functions for electrons in the atomic
ground state and first excited state which enter the definition of $\xi
_{coul}$ through $V_{\mathbf{R}}\left( _{\nu _{1}^{\prime }\ \nu _{1}}^{\nu
_{2}^{\prime }\ \nu _{2}}\right) $ defined in Eq. (2.21) are different, such
an electron-hole symmetry does not exist for Frenkel excitons. This makes
scattering like $\xi _{coul}\left( _{\mathbf{Q}_{1}^{\prime }\ \mathbf{Q}%
_{1}}^{\mathbf{Q}_{2}^{\prime }\ \mathbf{Q}_{2}}\right) $ different from
zero for $\mathbf{Q}_{1}^{\prime }=\mathbf{Q}_{1}$.

At this stage, we can roughly say that exchange Coulomb scatterings like the
ones of Fig. 2a are the relevant ones for Wannier excitons, as they are
responsible for the $\eta $ expansion of $\left\langle H\right\rangle _{N}$
while this type of scatterings reduces to zero for Frenkel excitons. In
contrast, diagrams like the ones of Fig. 2b are going to control Coulomb
exchange processes for Frenkel excitons while they reduce to zero for
Wannier excitons when one of the "in" excitons stay in the same state.

All this tends to show that fermion exchanges between Frenkel excitons are
unimportant in some configurations but important in a few others. So that a
careful calculation of these exchange processes through a secure many-body
theory as the one we have developed, is highly desirable to possibly trust
the obtained results.

\section{Hamiltonian mean value for $N=2$ excitons}

As for many other problems we have already tackled, it is quite valuable to
first perform calculations for $N=2$ because these calculations are far
simpler while most of the many-body physics already shows up.

In order to calculate $\left\langle H\right\rangle _{N}$ defined in Eq.
(1.2) for $N=2$, we first note that, due to Eqs. (1.3), (2.3)%
\begin{equation*}
\left\langle v\right\vert B_{\mathbf{0}}^{2}B_{\mathbf{0}}^{\dag
2}\left\vert v\right\rangle =2!F_{2}=2!\frac{N_{s}!}{(N_{s}-2)!N_{s}^{2}}
\end{equation*}%
\begin{equation}
=2\left( 1-\frac{1}{N_{s}}\right)   \tag{3.1}
\end{equation}%
while Eq. (2.4) taken for $N=2$, leads to%
\begin{equation}
HB_{\mathbf{0}}^{\dag 2}\left\vert v\right\rangle =2E_{\mathbf{0}}B_{\mathbf{%
0}}^{\dag 2}\left\vert v\right\rangle +\sum_{\mathbf{Q}}\xi \left( _{\mathbf{%
Q}\ \ \mathbf{0}}^{-\mathbf{Q}\ \ \mathbf{0}}\right) B_{\mathbf{Q}}^{\dag
}B_{-\mathbf{Q}}^{\dag }\left\vert v\right\rangle   \tag{3.2}
\end{equation}%
since $V_{\mathbf{0}}^{\dag }\left\vert v\right\rangle =0$ as easy to check
by taking Eq. (2.4) for $N=1$ acting on $\left\vert v\right\rangle $.The
Hamiltonian matrix element between two ground state excitons thus appears as%
\begin{equation}
\left\langle v\right\vert B_{\mathbf{0}}^{2}\left( H-2E_{\mathbf{0}}\right)
B_{\mathbf{0}}^{\dag 2}\left\vert v\right\rangle =\sum\limits_{\mathbf{Q}%
}\xi \left( _{\mathbf{Q}\ \ \ \mathbf{0}}^{\mathbf{-Q}\ \mathbf{0}}\right)
\left\langle v\right\vert B_{\mathbf{0}}^{2}B_{\mathbf{Q}}^{\dag }B_{-%
\mathbf{Q}}^{\dag }\left\vert v\right\rangle   \tag{3.3}
\end{equation}%
To go further, we use the scalar product of two Frenkel exciton states
which, according to Eqs. (2.1, 2, 9) reads as%
\begin{equation}
\left\langle v\right\vert B_{\mathbf{Q}_{1}^{\prime }}B_{\mathbf{Q}%
_{2}^{\prime }}B_{\mathbf{Q}_{2}}^{\dag }B_{\mathbf{Q}_{1}}^{\dag
}\left\vert v\right\rangle =\left( \delta _{\mathbf{Q}_{1}^{\prime }\text{ }%
\mathbf{Q}_{1}}\delta _{\mathbf{Q}_{2}^{\prime }\text{ }\mathbf{Q}%
_{2}}-\lambda \left( _{\mathbf{Q}_{1}^{\prime }\ \ \mathbf{Q}_{1}}^{\mathbf{Q%
}_{2}^{\prime }\ \ \mathbf{Q}_{2}}\right) \right) +(\mathbf{Q}_{1}^{\prime
}\leftrightarrow \mathbf{Q}_{2}^{\prime })  \tag{3.4}
\end{equation}%
When inserted into Eq. (3.3), the exchange Coulomb scattering generated by
the $\lambda $ term of this scalar product reduces to zero due to Eqs.
(2.24) and (2.26). We are thus left with%
\begin{equation}
\left\langle H\right\rangle _{2}=2E_{\mathbf{0}}+\frac{\xi \left( _{\mathbf{0%
}\ \mathbf{0}}^{\mathbf{0}\ \mathbf{0}}\right) }{1-1/N_{s}}  \tag{3.5}
\end{equation}

In view of this result, we are led to expect the Hamiltonian expectation
value for $N$ excitons to have a $N(N-1)/2$ prefactor in front of the
interaction term $\xi \left( _{\mathbf{0}\ \mathbf{0}}^{\mathbf{0}\ \mathbf{0%
}}\right) $\ of $\left\langle H\right\rangle _{2}$ since the two excitons
involved in this scattering are now taken among $N$; plus possibly higher
order terms in $N(N-1)(N-2)$..., due to fermion exchanges between more than
two excitons. In addition, the term $1/N_{s}$ in the denominator of Eq.
(3.5) could possibly be replaced by $(N-1)/N_{s}-\ldots
(N-1)(N-2)/N_{s}^{2}+...$ due again to fermion exchanges appearing in $F_{N}$%
.

We are going to show that the amount of $N$-body exchanges in the Coulomb
term exactly cancels the exchanges in $F_{N}$; so that the interaction term
of $\left\langle H\right\rangle _{N}$ ends by reading as the one for $%
\left\langle H\right\rangle _{2}$\ with just a $N(N-1)/2$ prefactor and
nothing else. Let us now prove this unexpected cancellation which actually
holds, within terms in $1/N_{s}$.

\section{Hamiltonian expectation value for $N$ Frenkel excitons}

According to Eq. (1.3), the denominator of $\left\langle H\right\rangle _{N}$%
\ in Eq. (1.2) is just $N!F_{N}$. To calculate the numerator, we use Eq.
(2.4) for $\left[ H,\text{ }B_{\mathbf{0}}^{\dag N}\right] $. This readily
gives%
\begin{equation*}
\left\langle H\right\rangle _{N}=NE_{\mathbf{0}}
\end{equation*}
\begin{equation}
+\frac{N(N-1)}{2}\sum_{\mathbf{Q}_{1}\mathbf{Q}_{2}}\xi \left( _{\mathbf{Q}%
_{1}\ \mathbf{0}}^{\mathbf{Q}_{2}\ \mathbf{0}}\right) \frac{\left\langle
v\right\vert B_{\mathbf{0}}^{N}B_{\mathbf{Q}_{1}}^{\dag }B_{\mathbf{Q}%
_{2}}^{\dag }B_{\mathbf{0}}^{\dag N-2}\left\vert v\right\rangle }{N!F_{N}} 
\tag{4.1}
\end{equation}

\subsection{Algebraic derivation}

Since scattering conserves momentum, $\xi \left( _{\mathbf{Q}_{1}\ \mathbf{0}%
}^{\mathbf{Q}_{2}\ \mathbf{0}}\right) $ differs from zero for $\mathbf{Q}%
_{1}+\mathbf{Q}_{2}=0$ only. So that to get $\left\langle H\right\rangle _{N}
$, we must calculate%
\begin{equation}
G_{N}(\mathbf{Q})=\left\langle v\right\vert B_{\mathbf{0}}^{N}B_{\mathbf{Q}%
}^{\dag }B_{-\mathbf{Q}}^{\dag }B_{\mathbf{0}}^{\dag N-2}\left\vert
v\right\rangle   \tag{4.2}
\end{equation}%
As for $\mathbf{Q}=\mathbf{0}$, this scalar product is just $N!F_{N}$, we in
fact need to calculate $G_{N}(\mathbf{Q})$ for $\mathbf{Q\neq 0}$ only.
Since $D_{\mathbf{Q}0}\left\vert v\right\rangle =0$ as seen from Eq. (2.12)
taken for $N=1$ acting on vacuum, this Eq. (2.12) readily gives%
\begin{equation}
G_{N}^{\ast }(\mathbf{Q\neq 0})=-\frac{N(N-1)}{N_{s}}\left\langle
v\right\vert B_{\mathbf{0}}^{N-2}B_{\mathbf{Q}}B_{\mathbf{Q}}^{\dag }B_{%
\mathbf{0}}^{\dag N-2}\left\vert v\right\rangle   \tag{4.3}
\end{equation}%
To go further, we use $\left[ B_{\mathbf{Q}},B_{\mathbf{Q}}^{\dagger }\right]
=1-D_{\mathbf{QQ}}$. This leads to split the RHS of the above equation as%
\begin{equation*}
G_{N}^{\ast }(\mathbf{Q}\mathbf{\neq 0})=-\frac{N(N-1)}{N_{s}}\left[
\left\langle v\right\vert B_{\mathbf{0}}^{N-2}B_{\mathbf{0}}^{\dag
N-2}\left\vert v\right\rangle \right. 
\end{equation*}%
\begin{equation*}
-\left\langle v\right\vert B_{\mathbf{0}}^{N-2}D_{\mathbf{QQ}}B_{\mathbf{0}%
}^{\dag N-2}\left\vert v\right\rangle 
\end{equation*}
\begin{equation}
\left. +\left\langle v\right\vert B_{\mathbf{0}}^{N-2}B_{\mathbf{Q}}^{\dag
}B_{\mathbf{Q}}B_{\mathbf{0}}^{\dag N-2}\left\vert v\right\rangle \right]  
\tag{4.4}
\end{equation}%
The first term is just $(N-2)!F_{N-2}$. The second term, calculated from Eq.
(2.11), gives $\left[ 2(N-2)/N_{s}\right] (N-2)!F_{N-2}$. In the last term,
we use Eq. (2.12) for $B_{\mathbf{Q}}B_{\mathbf{0}}^{\dag N-2}\left\vert
v\right\rangle $. This allows us to rewrite this term as $-\left[
(N-2)(N-3)/N_{s}\right] G_{N-2}(\mathbf{Q})$. By collecting these three
terms, we get a recursion relation \ between the $G_{N}$'s. It reads%
\begin{equation*}
G_{N}^{\ast }(\mathbf{Q}\mathbf{\neq 0})=-\frac{N(N-1)}{N_{s}}\left( 1-2%
\frac{N-2}{N_{s}}\right) (N-2)!F_{N-2}
\end{equation*}
\begin{equation}
+\frac{N(N-1)(N-2)(N-3)}{N_{s}^{2}}G_{N-2}(\mathbf{Q})  \tag{4.5}
\end{equation}%
Using Eq. (2.13) for $F_{N}$, it is then easy to check that this recursion
relation is fulfilled for 
\begin{equation}
G_{N}(\mathbf{Q}\mathbf{\neq 0})=-\frac{1}{N_{s}-1}N!F_{N}  \tag{4.6}
\end{equation}%
So that we end with a surprisingly simple expression for $G_{N}(\mathbf{Q})$%
, namely%
\begin{equation}
G_{N}(\mathbf{Q})=\frac{N_{s}\delta _{\mathbf{Q}\text{ }\mathbf{0}}-1}{%
N_{s}-1}N!F_{N}  \tag{4.7}
\end{equation}%
When inserted into the sum of Eq. (4.1), we find that this sum reduces to%
\begin{equation}
\sum_{\mathbf{Q}}\xi \left( _{\mathbf{Q}\ \mathbf{0}}^{-\mathbf{Q}\ \mathbf{0%
}}\right) \frac{N_{s}\delta _{\mathbf{Q}\text{ }\mathbf{0}}-1}{N_{s}-1} 
\tag{4.8}
\end{equation}%
Since the sum of Coulomb scatterings over $\mathbf{Q}$\ gives zero according
to Eqs. (2.25, 26), we get the following compact form for the Hamiltonian
mean value of $N$ Frenkel excitons%
\begin{equation}
\left\langle H\right\rangle _{N}=NE_{0}+\frac{N(N-1)}{2}\frac{\xi \left( _{%
\mathbf{0}\ \mathbf{0}}^{\mathbf{0}\ \mathbf{0}}\right) }{1-1/N_{s}} 
\tag{4.9}
\end{equation}%
It shows that the interaction term is indeed the one for two Frenkel
excitons with just a $N(N-1)/2$ prefactor which corresponds to the number of
ways to choose these 2 excitons among $N$.

\subsection{Shiva diagram derivation}

Although the above algebraic derivation is quite easy to follow, it may
leave the reader unsatisfied as it does not really show why the amount of
exchange processes in the Coulomb term is exactly the same as the one
appearing in the normalization factor within terms in $1/N_{s}$. To grasp
this reason, Shiva diagrams once more are quite valuable.

Let us consider the Coulomb term appearing in $\left\langle H\right\rangle
_{N}$, namely%
\begin{equation}
C_{N}=-\frac{N(N-1)}{N_{s}}\sum_{\mathbf{Q}_{1}\mathbf{Q}_{2}}\left\langle
v\right\vert B_{\mathbf{0}}^{N}B_{\mathbf{Q}_{1}}^{\dag }B_{\mathbf{Q}%
_{2}}^{\dag }B_{\mathbf{0}}^{\dag N-2}\left\vert v\right\rangle \xi \left( _{%
\mathbf{Q}_{1}\ \mathbf{0}}^{\mathbf{Q}_{2}\ \mathbf{0}}\right)  \tag{4.10}
\end{equation}%
In it, two excitons $\mathbf{0}$ have a direct Coulomb interaction to become
($\mathbf{Q}_{1}$, $\mathbf{Q}_{2}$). They can then possibly mix with the
other $(N-2)$ excitons $\mathbf{0}$\ through fermion exchanges to end as $N$
excitons $\mathbf{0}$. This sum thus corresponds to to the diagram of Fig.
3a.

The simplest of these exchanges corresponds to leave $(N-1)$ excitons $%
\mathbf{0}$ unaffected. So that these $(N-2)$ excitons appear as $%
\left\langle v\right\vert B_{\mathbf{0}}^{N-2}B_{\mathbf{0}}^{\dag
N-2}\left\vert v\right\rangle =(N-2)!F_{N-2}$. The two other excitons ($%
\mathbf{Q}_{1}$, $\mathbf{Q}_{2}$) become ($\mathbf{0}$, $\mathbf{0}$) with
or without exchanging their fermions. This first possibility which
corresponds to the diagram of Fig. 3b, thus leads to 
\begin{equation}
\left[ (N-2)!F_{N-2}\right] \left[ N(N-1)\left( \xi \left( _{\mathbf{0}\ 
\mathbf{0}}^{\mathbf{0}\ \mathbf{0}}\right) -\xi ^{in}\left( _{\mathbf{0}\ 
\mathbf{0}}^{\mathbf{0}\ \mathbf{0}}\right) \right) \right]  \tag{4.11}
\end{equation}%
the $N(N-1)$ factor in the second bracket comes from the number of ways to
choose the two excitons $\mathbf{0}$ on the left among $N$.

A second possibility is to mix one exciton $\mathbf{0}$ with the excitons ($%
\mathbf{Q}_{1}$, $\mathbf{Q}_{2}$)\ while leaving the other $(N-3)$ excitons 
$\mathbf{0}$\ apart. This second possibility corresponds to the diagram of
Fig. 3c. We have shown in Eq. (2.27) that Coulomb exchange scatterings like
the one of Fig. 2a reduce to zero. By using the value of the exchange Shiva
diagram of Fig. 1b given in Eq. (2.10), we then find from the diagram of
Fig. 3c that this second possibility produces a term%
\begin{equation}
-\left[ (N-3)!F_{N-3}\right] \left[ N(N-1)(N-2)\frac{1}{N_{s}}\xi \left( _{%
\mathbf{0}\ \mathbf{0}}^{\mathbf{0}\ \mathbf{0}}\right) (N-2)\right] 2 
\tag{4.12}
\end{equation}%
in the Shiva expansion of $C_{N}$. The factor $N(N-1)(N-2)$ comes from the
number of ways to choose the three excitons $\mathbf{0}$ of the left among $%
N $ while the factor $(N-2)$ comes from the number of ways to choose the
exciton $\mathbf{0}$ on the right among $(N-2)$. This term has an overall
minus sign because one fermion exchange is involved. An additional factor 2
comes from the number of diagrams which give the same nonzero contributions
and which have the structure like the diagram of Fig. 2b.

A third possibility is to mix two excitons $\mathbf{0}$ with the excitons ($%
\mathbf{Q}_{1}$, $\mathbf{Q}_{2}$). The nonzero Coulomb exchange diagrams
are shown in Fig. 4. They leads to%
\begin{equation}
+\left[ (N-4)!F_{N-4}\right] \left[ N(N-1)(N-2)(N-3)\frac{1}{N_{s}^{2}}\xi
\left( _{\mathbf{0}\ \mathbf{0}}^{\mathbf{0}\ \mathbf{0}}\right) (N-2)(N-3)%
\right] 3  \tag{4.13}
\end{equation}%
and so on...

This Shiva diagram expansion of the Coulomb term $C_{N}$ defined in Eq.
(4.10) allows us to write it as%
\begin{equation}
C_{N}=S_{N}\xi \left( _{\mathbf{0}\ \mathbf{0}}^{\mathbf{0}\ \mathbf{0}%
}\right)   \tag{4.14}
\end{equation}%
\begin{equation*}
S_{N}=(N-2)!F_{N-2}N(N-1)-(N-3)!F_{N-3}N(N-1)(N-2)\frac{2}{N_{s}}(N-2)
\end{equation*}%
\begin{equation}
+(N-4)!F_{N-4}N(N-1)(N-2)(N-3)\frac{3}{N_{s}^{2}}\xi \left( _{\mathbf{0}\ 
\mathbf{0}}^{\mathbf{0}\ \mathbf{0}}\right) (N-2)(N-3)-...  \tag{4.15}
\end{equation}

The last step is to relate $S_{N}$ to $F_{N}$. A simple way to do it is to
consider the scalar product%
\begin{equation}
P_{N}(\mathbf{Q}_{1},\mathbf{Q}_{2})=\left\langle v\right\vert B_{\mathbf{0}%
}^{N}B_{\mathbf{Q}_{1}}^{\dag }B_{\mathbf{Q}_{2}}^{\dag }B_{\mathbf{0}%
}^{\dag N-2}\left\vert v\right\rangle   \tag{4.16}
\end{equation}%
shown in Fig. 5a. Let us perform a Shiva diagram expansion of this scalar
product according to the standard rules \cite{EPJB}, namely we isolate $(N-2)
$, $(N-3)$, ... excitons $\mathbf{0}$ and write all the possible diagrams in
which the remaining excitons $\mathbf{0}$ are "not alone". This leads to the
diagram of Fig. 5c for the second term. And so on...: By using the results
for Shiva diagrams in the case of Frenkel excitons given in Eqs. (2.9, 10),
we get for $\mathbf{Q}_{1}=\mathbf{Q}_{2}=0$, i.e., when $P_{N}(\mathbf{Q}%
_{1},\mathbf{Q}_{2})$ reduces to $N!F_{N}$%
\begin{equation*}
N!F_{N}=(N-2)!F_{N-2}N(N-1)\left( 1-\frac{1}{N_{s}}\right) 
\end{equation*}%
\begin{equation*}
+(N-3)!F_{N-3}N(N-1)(N-2)\left( \frac{2}{N_{s}}-\frac{2}{N_{s}^{2}}\right)
(N-2)+...
\end{equation*}
\begin{equation}
=\left( 1-\frac{1}{N_{s}}\right) S_{N}  \tag{4.17}
\end{equation}

When inserted into Eqs. (4.14) and (4.1), this result allows one to readily
recover the expression of the Hamiltonian mean value for $N$ Frenkel
excitons given in Eq. (4.9).

It is clear that such a derivation needs to be knowledgeable with Shiva
diagrams. It however has the great advantage to pick out in a transparent
way from where the final result comes. In the present case, this is from
Coulomb exchange processes like the one of Fig. 2.

\subsection{Physical understanding}

In order to identify the microscopic Coulomb processes which control the
ground state energy of $N$ Frenkel excitons, we first use the definition of
the interaction scattering given in Eqs. (2.15, 16, 19). This leads to%
\begin{equation}
\left\langle H\right\rangle _{N}=NE_{0}+\frac{N(N-1)}{2(N_{s}-1)}(\mathcal{W}%
_{\mathbf{0}}-2\mathcal{V}_{\mathbf{0}})  \tag{4.18}
\end{equation}%
When compared to Eq. (1.4), the microscopic value of the Coulomb
contribution to the energy is thus found to read $\gamma _{\mathbf{0}}=%
\mathcal{W}_{\mathbf{0}}-2\mathcal{V}_{\mathbf{0}}$.

To get a precise understanding of the various contributions to the
Hamiltonian mean value, let us now write the above quantities in terms of
the various elementary Coulomb scatterings between atomic sites. For that,
we use Eqs. (2.16, 17, 18, 19, 20) and the expression of the $\mathbf{Q}$
exciton energy, namely $E_{\mathbf{Q}}=E_{pair}+\mathcal{V}_{\mathbf{Q}}$
where the free pair energy $E_{pair}=\varepsilon _{e}+\varepsilon
_{h}-\varepsilon _{0}$ with $\varepsilon _{0}=V_{\mathbf{0}}\left( _{0\
0}^{1\ 1}\right) -V_{\mathbf{0}}\left( _{1\ 0}^{0\ 1}\right) $ is just the
energy of one electron in the excited level and one hole in the ground state
level, minus the neutrality energy $\varepsilon _{0}$ we would pay for not
having electron and hole on the same atomic site (see Ref. \cite{paper1} ).
This leads to%
\begin{equation*}
\left\langle H\right\rangle _{N}=N\left[ \varepsilon _{e}+\varepsilon
_{h}-V_{\mathbf{0}}\left( _{0\ 0}^{1\ 1}\right) +\sum_{\mathbf{R}}V_{\mathbf{%
R}}\left( _{1\ 0}^{0\ 1}\right) \right. 
\end{equation*}%
\begin{equation}
\left. +\frac{(N-1)}{2(N_{s}-1)}\sum_{\mathbf{R\neq 0}}\left[ V_{\mathbf{R}%
}\left( _{1\ 1}^{1\ 1}\right) +V_{\mathbf{R}}\left( _{0\ 0}^{0\ 0}\right)
-V_{\mathbf{R}}\left( _{0\ 0}^{1\ 1}\right) -V_{\mathbf{R}}\left( _{1\
1}^{0\ 0}\right) -V_{\mathbf{R}}\left( _{1\ 0}^{0\ 1}\right) -V_{\mathbf{R}%
}\left( _{0\ 1}^{1\ 0}\right) \right] \right]   \tag{4.19}
\end{equation}%
since in the last sum, we can replace $2V_{\mathbf{R}}\left( _{1\ 0}^{0\
1}\right) $ by $V_{\mathbf{R}}\left( _{1\ 0}^{0\ 1}\right) +V_{\mathbf{R}%
}\left( _{0\ 1}^{1\ 0}\right) $, due to Eq. (2.21).

We see that the interaction term, which cancels for $N=1$, as expected,
contains all possible Coulomb processes between the atomic levels 0 and 1 of
the different atomic sites, through the sum over $\mathbf{R\neq 0}$. Due to
the orhogonality of the $\nu =(0,1)$ atomic states, it is of importance to
note that the indirect Coulomb processes $V_{\mathbf{R}}\left( _{1\ 0}^{0\
1}\right) $ in which electrons change from atomic state 1 to atomic state 0
are expected to be much smaller than the direct Coulomb processes $V_{%
\mathbf{R}}\left( _{0\ 0}^{1\ 1}\right) $; so that the interaction term of
the Hamiltonian expectation value is essentially controlled by direct
Coulomb processes $V_{\mathbf{R}}\left( _{\nu _{1}^{\prime }\ \nu _{1}}^{\nu
_{2}^{\prime }\ \nu _{2}}\right) $. Let us however stress that, while these
indirect scatterings are not that important in the interaction term of the
Hamiltonian mean value, they play a major role in the Frenkel exciton
physics because, although extremely small, they are the only processes
allowing an excitation transfer between sites as necessary to produce the
exciton.

Since the Coulomb matrix elements $V_{\mathbf{R}}\left( _{\nu _{1}^{\prime
}\ \nu _{1}}^{\nu _{2}^{\prime }\ \nu _{2}}\right) $ entering the
Hamiltonian mean value are defined in terms of wave functions for electron
in the ground and first excited atomic states, they strongly depend on the
chemical properties of the material at hand. This has to be contrasted with
Wannier excitons which are constructed on free electrons and free holes, the
material of interest just appearing through effective masses induced by the
lattice periodicity. For these Wannier excitons, the intraband Coulomb
matrix element thus reduces to $4\pi e^{2}/L^{3}q^{2}$ in 3D. As a result,
the Hamiltonian mean value $\left\langle H\right\rangle _{N}$ for these
excitons only depends on one input parameter which is the exciton Bohr
radius; this is why its $\eta $\ expansion can be derived explicitly for all
materials (see Eq. (1.6)). In contrast, the prefactor of the $\eta $\ term
for Frenkel excitons highly depends on the material at hand.

If we now come back to Eq. (4.18), we can, in the large sample limit,
replace $N_{s}-1$ by $N_{s}$. So that, for $N$ large compared to 1, the
Hamiltonian expectation value expands as given in Eq. (1.5): It only
contains a linear term in density. Again, this has to be contrasted with $%
\left\langle H\right\rangle _{N}$ for Wannier excitons in which the Coulomb
term reads as a density expansion in $\eta $.

For completeness, let us note that the Hamiltonian mean value for $N$
identical elementary bosons also has just one linear contribution in
density. This is expectable since the Hamiltonian mean value contains the
interaction at first order only by construction: when the particles are
taken as elementary bosons the $2\times 2$ interaction couples two bosons
among $N$, no fermion exchange exists to possibly mix these two particles
with the rest of the system to produce higher order couplings. This is
supported by the precise calculation of the Hamiltonian mean value%
\begin{equation}
\left\langle \overline{H}\right\rangle _{N}=\frac{\left\langle v\right\vert 
\overline{B}_{\mathbf{0}}^{N}\overline{H}\overline{B}_{\mathbf{0}}^{\dag
N}\left\vert v\right\rangle }{\left\langle v\right\vert \overline{B}_{%
\mathbf{0}}^{N}\overline{B}_{\mathbf{0}}^{\dag N}\left\vert v\right\rangle }
\tag{4.20}
\end{equation}%
which for bosonized excitons such that $\left[ \overline{B}_{\mathbf{Q}%
^{\prime }},\overline{B}_{\mathbf{Q}}^{\dag }\right] =\delta _{\mathbf{Q}%
^{\prime }\mathbf{Q}}$, has a denominator which reduces to $N!$ while, for
an effective bosonic Hamiltonian written as%
\begin{equation}
\overline{H}=\sum E_{\mathbf{Q}}\overline{B}_{\mathbf{Q}}^{\dag }\overline{B}%
_{\mathbf{Q}}+\frac{1}{2}\sum \overline{V}\left( _{\mathbf{Q}_{1}^{\prime }\ 
\mathbf{Q}_{1}}^{\mathbf{Q}_{2}^{\prime }\ \mathbf{Q}_{2}}\right) \overline{B%
}_{\mathbf{Q}_{1}^{\prime }}^{\dag }\overline{B}_{\mathbf{Q}_{2}^{\prime
}}^{\dag }\overline{B}_{\mathbf{Q}_{2}}\overline{B}_{\mathbf{Q}_{1}} 
\tag{4.21}
\end{equation}%
its numerator follows, since $\overline{B}_{\mathbf{Q}}\overline{B}_{\mathbf{%
0}}^{\dag N}\left\vert v\right\rangle =N\delta _{\mathbf{Q}\text{ }\mathbf{0}%
}\overline{B}_{\mathbf{0}}^{\dag N-1}\left\vert v\right\rangle $, from%
\begin{equation*}
\left\langle v\right\vert \overline{B}_{\mathbf{0}}^{N}\left( \overline{H}%
-NE_{\mathbf{0}}\right) \overline{B}_{\mathbf{0}}^{\dag N}\left\vert
v\right\rangle =\frac{N(N-1)}{2}\sum \overline{V}\left( _{\mathbf{Q}%
_{1}^{\prime }\ \mathbf{0}}^{\mathbf{Q}_{2}^{\prime }\ \mathbf{0}}\right) 
\end{equation*}%
\begin{equation}
\left\langle v\right\vert \overline{B}_{\mathbf{0}}^{N}\overline{B}_{\mathbf{%
Q}_{1}^{\prime }}^{\dag }\overline{B}_{\mathbf{Q}_{2}^{\prime }}^{\dag }%
\overline{B}_{\mathbf{0}}^{\dag N-2}\left\vert v\right\rangle   \tag{4.22}
\end{equation}%
This leads to an Hamiltonian mean value given by%
\begin{equation}
\left\langle \overline{H}\right\rangle _{N}=NE_{\mathbf{0}}+\frac{N(N-1)}{2}%
\overline{V}\left( _{\mathbf{0}\ \mathbf{0}}^{\mathbf{0}\ \mathbf{0}}\right) 
\tag{4.23}
\end{equation}%
By noting that the scattering $\overline{V}\left( _{\mathbf{0}\ \mathbf{0}}^{%
\mathbf{0}\ \mathbf{0}}\right) $ must decrease as $1/L^{D}$ with the sample
volume, as necessary for the extensivity of the system, we see that the
correction to the energy of one bosonized exciton just increases linearly
with the exciton density $N/L^{D}$.

\section{Conclusion}

In this paper, we calculate the ground state energy of $N$ Frenkel excitons
in the Born approximation through the Hamiltonian mean value in the state
made of $N$ ground state excitons. We show that the energy change induced by
exciton interactions contains one linear term only in the exciton density,
while higher order terms appear in the case of Wannier excitons. A naive
reason could be that, since Frenkel excitons are constructed on highly
localized atomic wave functions, their exchange Coulomb scatterings must
reduce to zero, making Frenkel excitons appearing as elementary bosons. This
however does not hold since the normalization factor of $N$ Frenkel excitons
is exponentially small compared to the one for $N$ elementary bosons. The
deep nonobvious reason for having one linear term only in the Hamiltonian
mean value, is the fact that the same exponentially small prefactor also
exists in the Coulomb term of the Hamiltonian matrix element, within
corrections in $1/N_{s}$ where $N_{s}$\ is the number of atomic sites. So
that, in the end, all the exchange processes appearing in $\left\langle
H\right\rangle _{N}$ finally disappear, as if Frenkel excitons were true
elementary particles, within terms in $1/N_{s}$. Here again, Shiva diagrams
enlighten the physical understanding of the fermion exchanges which lead to
this rather unexpected cancellation.

\acknowledgments

W. V. P. is supported by the Ministry of Education of France, the Russian
Science Support Foundation, and the President of Russia program for young
scientists.

\begin{figure}
\begin{center}
\centering
   \subfigure{\includegraphics[width=0.3\textwidth]{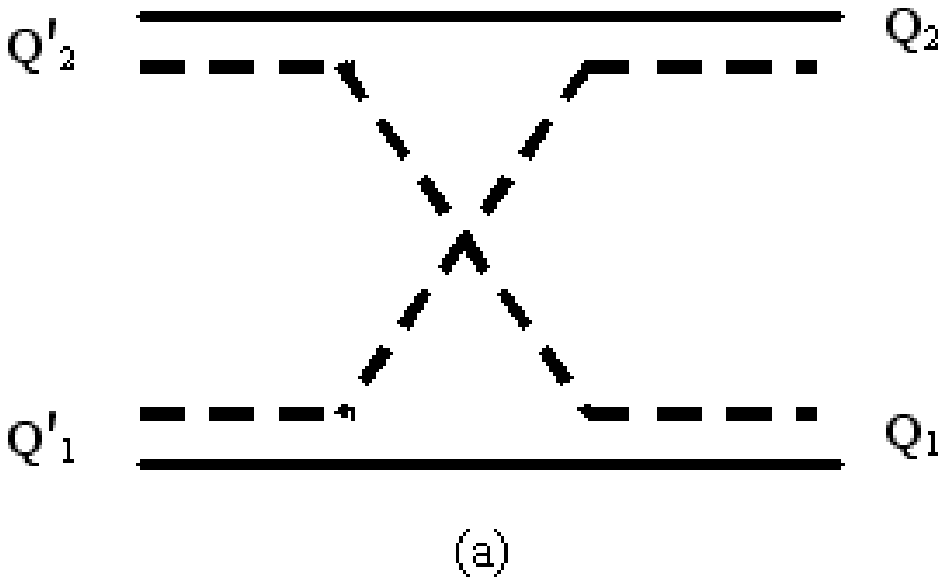}}\\
   \subfigure{\includegraphics[width=0.3\textwidth]{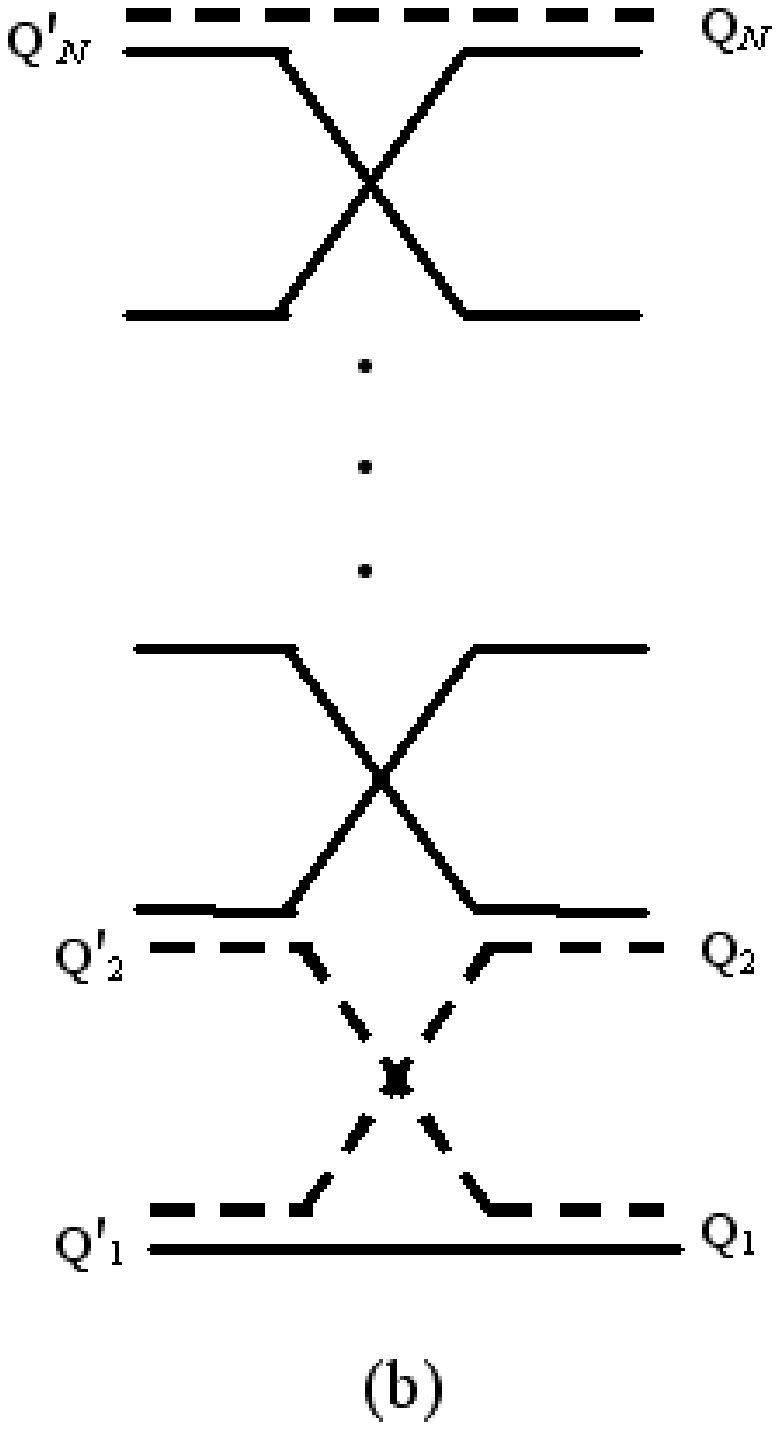}}
\end{center} \caption{\label{Fig1}
(a) Shiva diagrams for carrier exchanges between two Frenkel excitons. This
diagram represents the Pauli scattering $\lambda \left( _{\mathbf{Q}%
_{1}^{\prime }\ \mathbf{Q}_{1}}^{\mathbf{Q}_{2}^{\prime }\ \mathbf{Q}%
_{2}}\right) $ given in Eq. (2.9).
(b) Shiva diagram for carrier exchanges between $N$ Frenkel excitons, as
given in Eq. (2.10).}
\end{figure}

\begin{figure}
\begin{center}
\centering
   \subfigure{\includegraphics[width=0.3\textwidth]{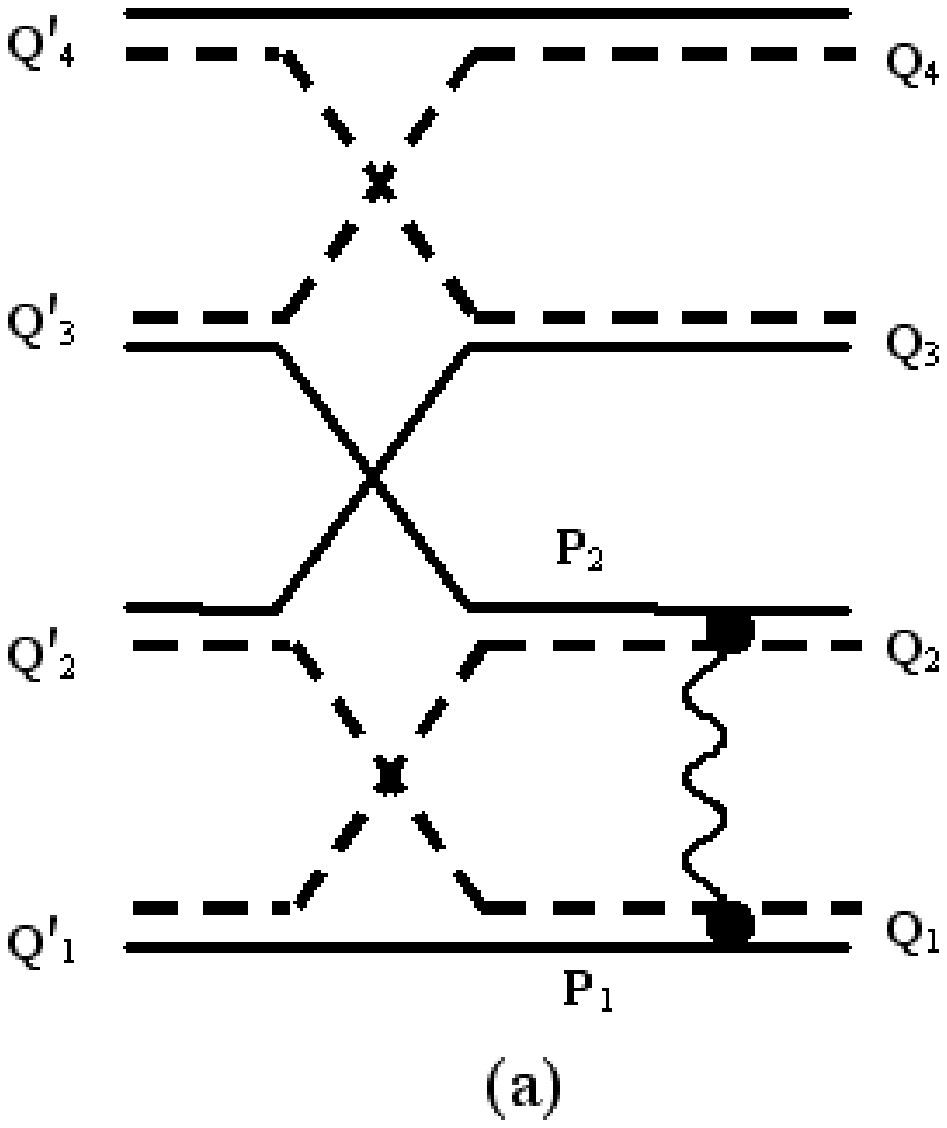}}\\
   \subfigure{\includegraphics[width=0.3\textwidth]{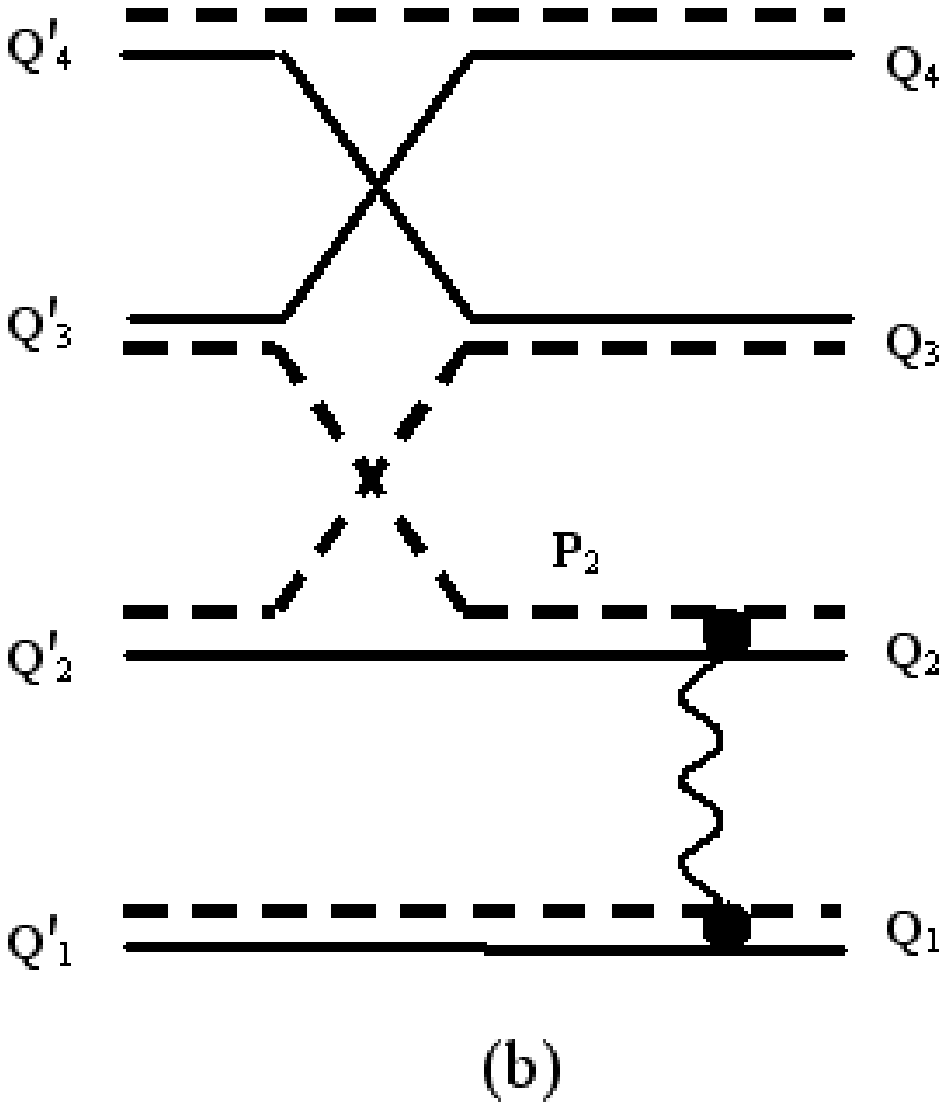}}\\
   \subfigure{\includegraphics[width=0.3\textwidth]{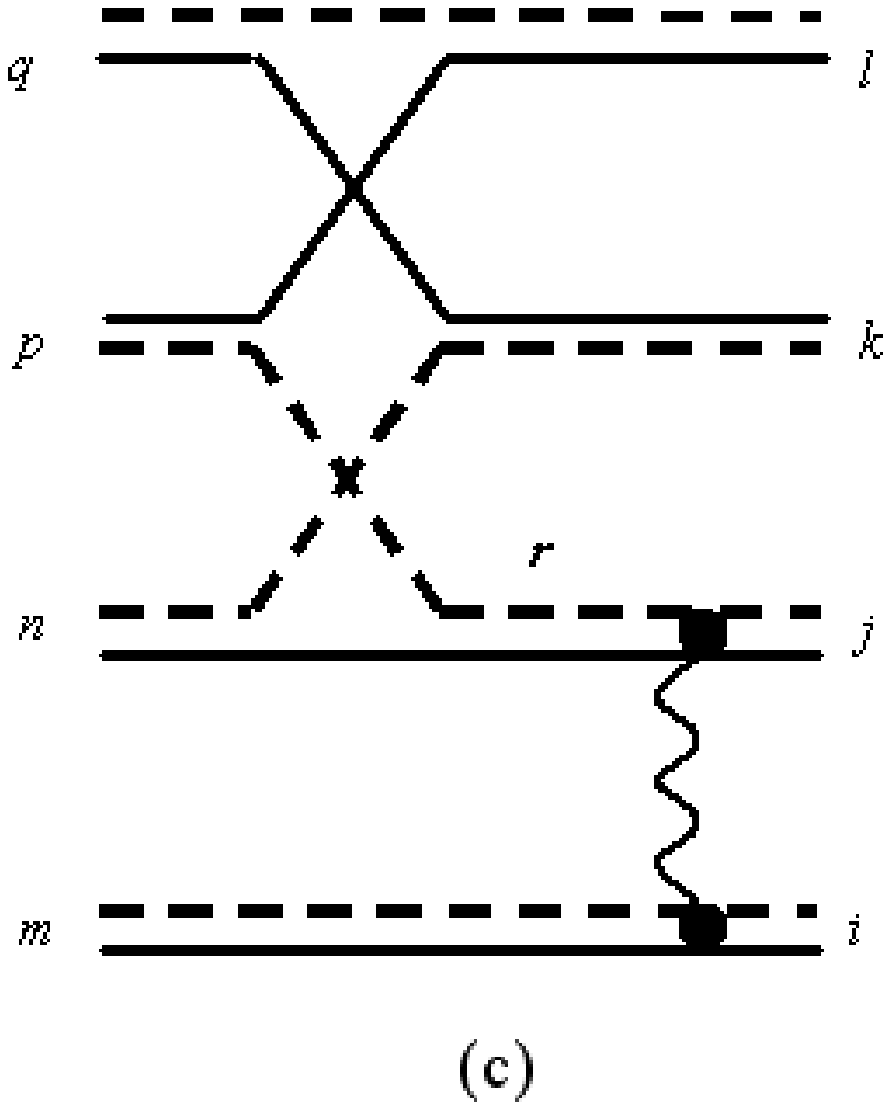}}
\end{center} \caption{\label{Fig2}
(a) Coulomb exchange processes between 4 Frenkel excitons in which all "in"
excitons and "out" excitons are made with different electron-hole pairs.
This type of processes reduces to zero for Frenkel excitons (see Eq. (2.27)).
(b) Coulomb exchange process in which one Frenkel exciton keeps its carrier
pair (see Eq. (2.28)).
(c) Same as (b) for Wannier excitons characterized by $i=(\mathbf{Q}_{1},\nu
_{1})$. This type of Coulomb exchange process reduces to zero for Wannier
excitons when the exciton which keeps its fermion stays in the same state ($%
m=i$).}
\end{figure}

\begin{figure}
\begin{center}
\centering
   \subfigure{\includegraphics[width=0.4\textwidth]{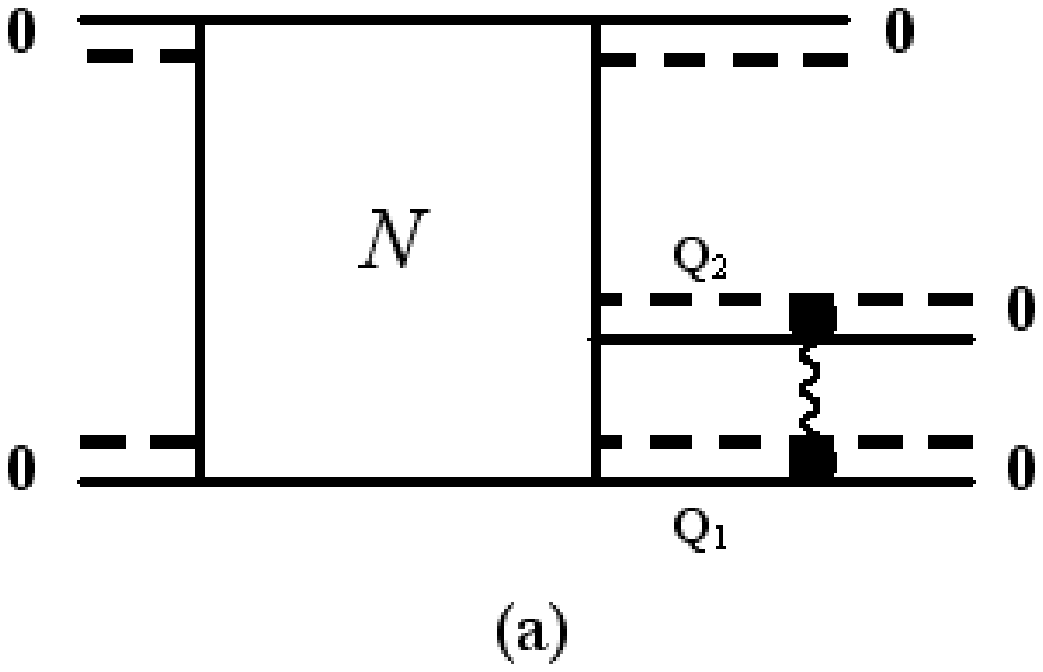}}\\
   \subfigure{\includegraphics[width=0.6\textwidth]{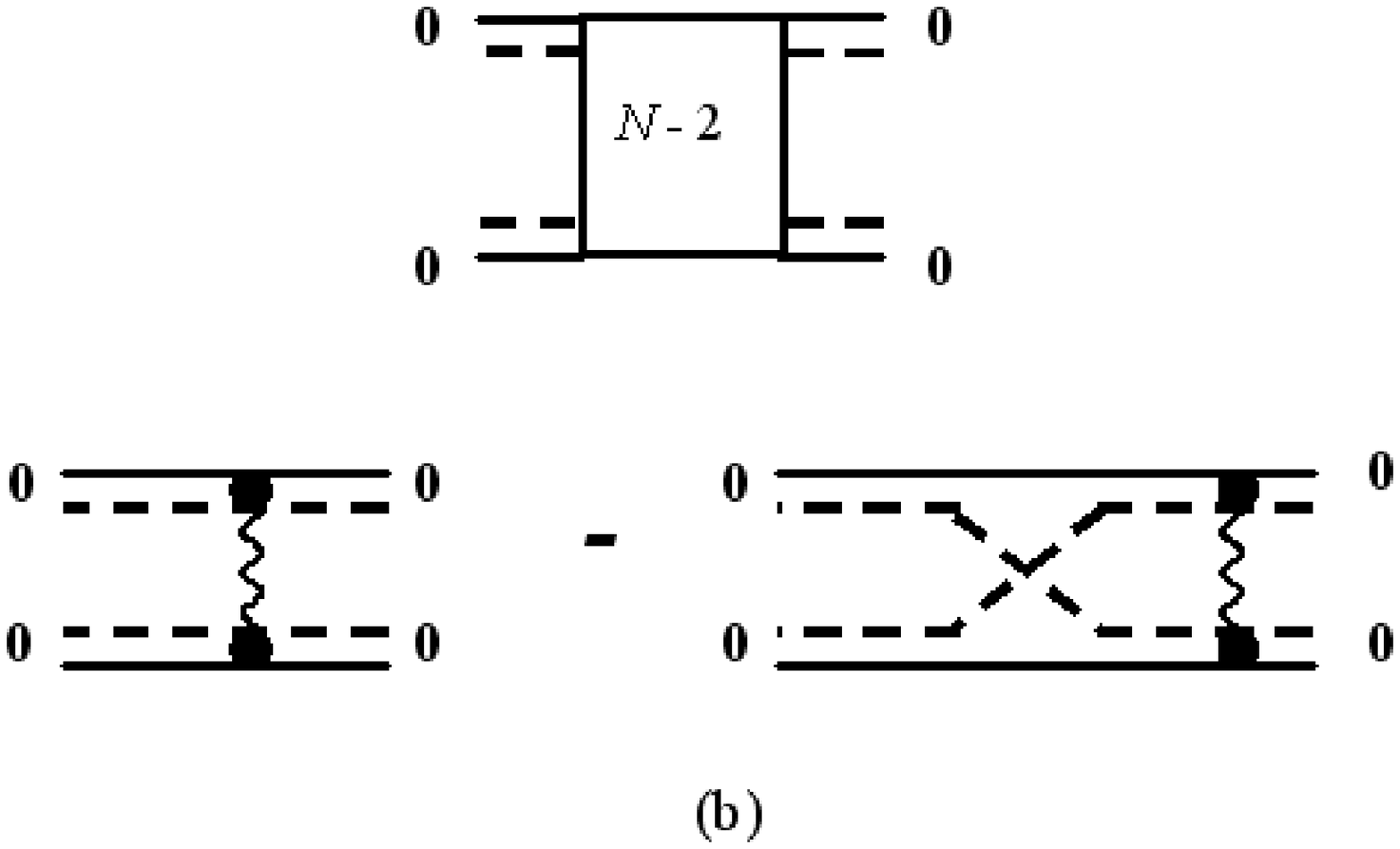}}\\
   \subfigure{\includegraphics[width=0.6\textwidth]{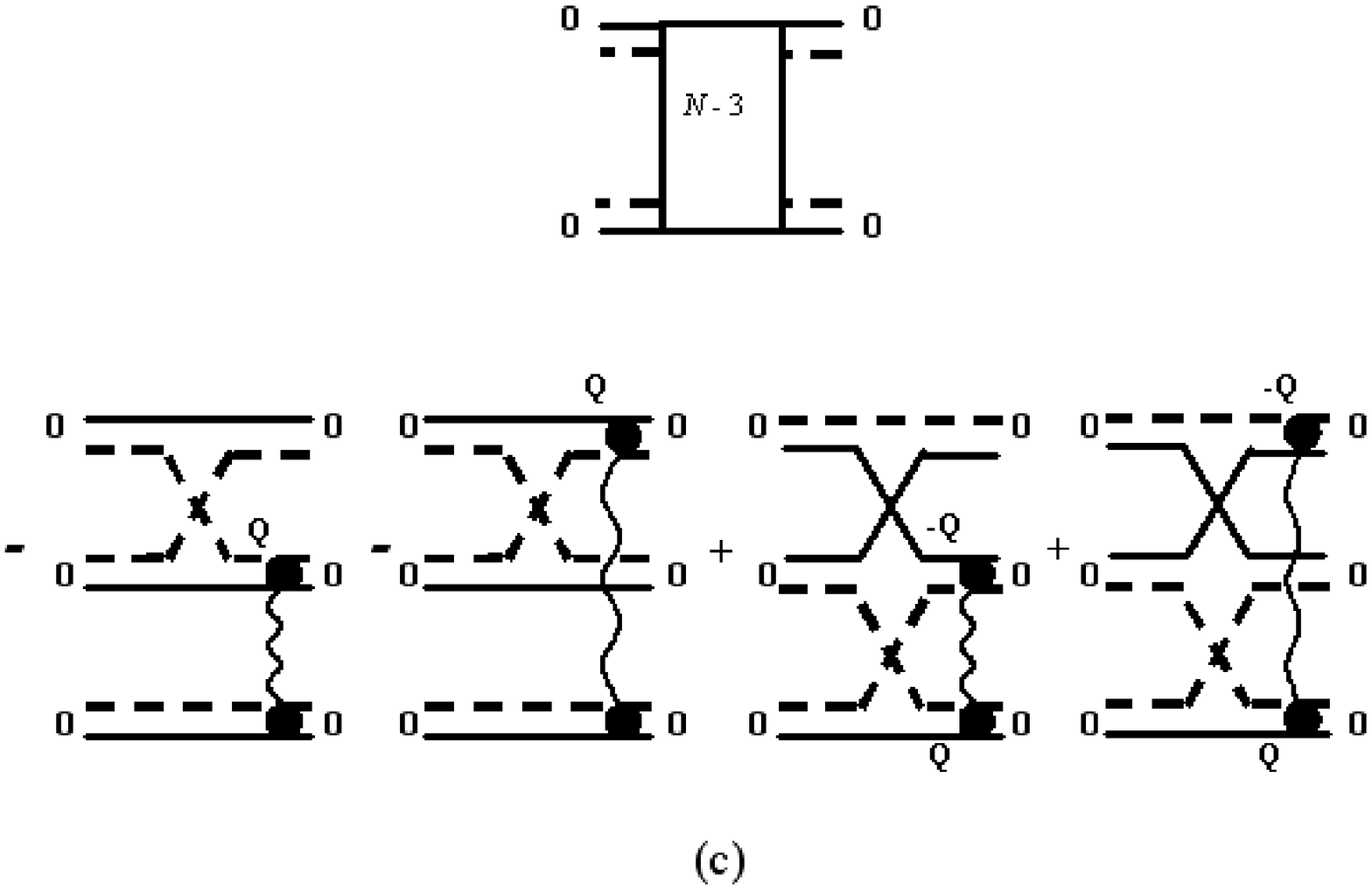}}
\end{center} \caption{\label{Fig3}
(a) Shiva diagram representation of the Coulomb part of the Hamiltonian mean
value, as given in Eq. (4.9).
(b) First order term: $(N-2)$ excitons $\mathbf{0}$ stay unaffected in this
process.
(c) Second order term: one exciton $\mathbf{0}$ among $(N-2)$ exchanges its
fermion with the ones scattered by the $2\times 2$ Coulomb interaction.}
\end{figure}

\begin{figure} \begin{center}
\includegraphics[width=0.4\textwidth]{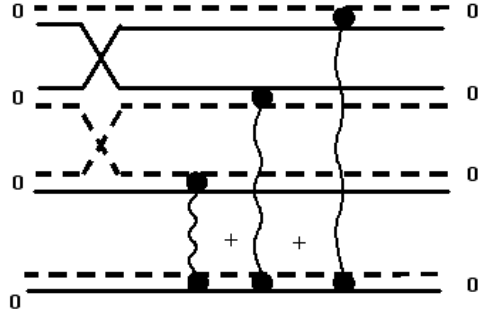}\end{center}
\caption{\label{Fig4} Non-zero Coulomb exchange process in which two excitons $\mathbf{0}$ among $%
(N-2)$ exchange their fermions with the ones scattered by the $2\times 2$\
Coulomb interaction.}
\end{figure}

\begin{figure}
\begin{center}
\centering
   \subfigure{\includegraphics[width=0.4\textwidth]{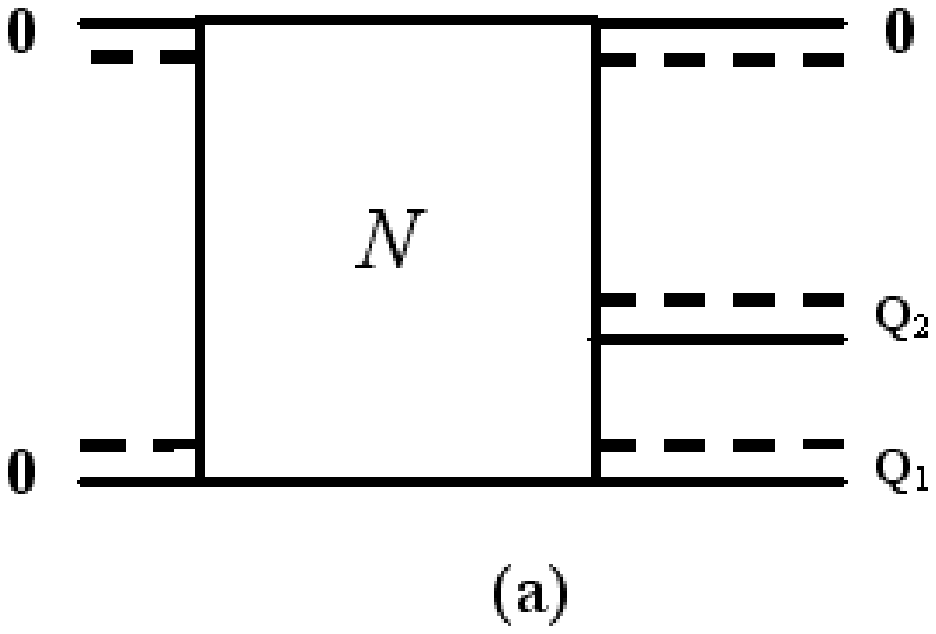}}\\
   \subfigure{\includegraphics[width=0.5\textwidth]{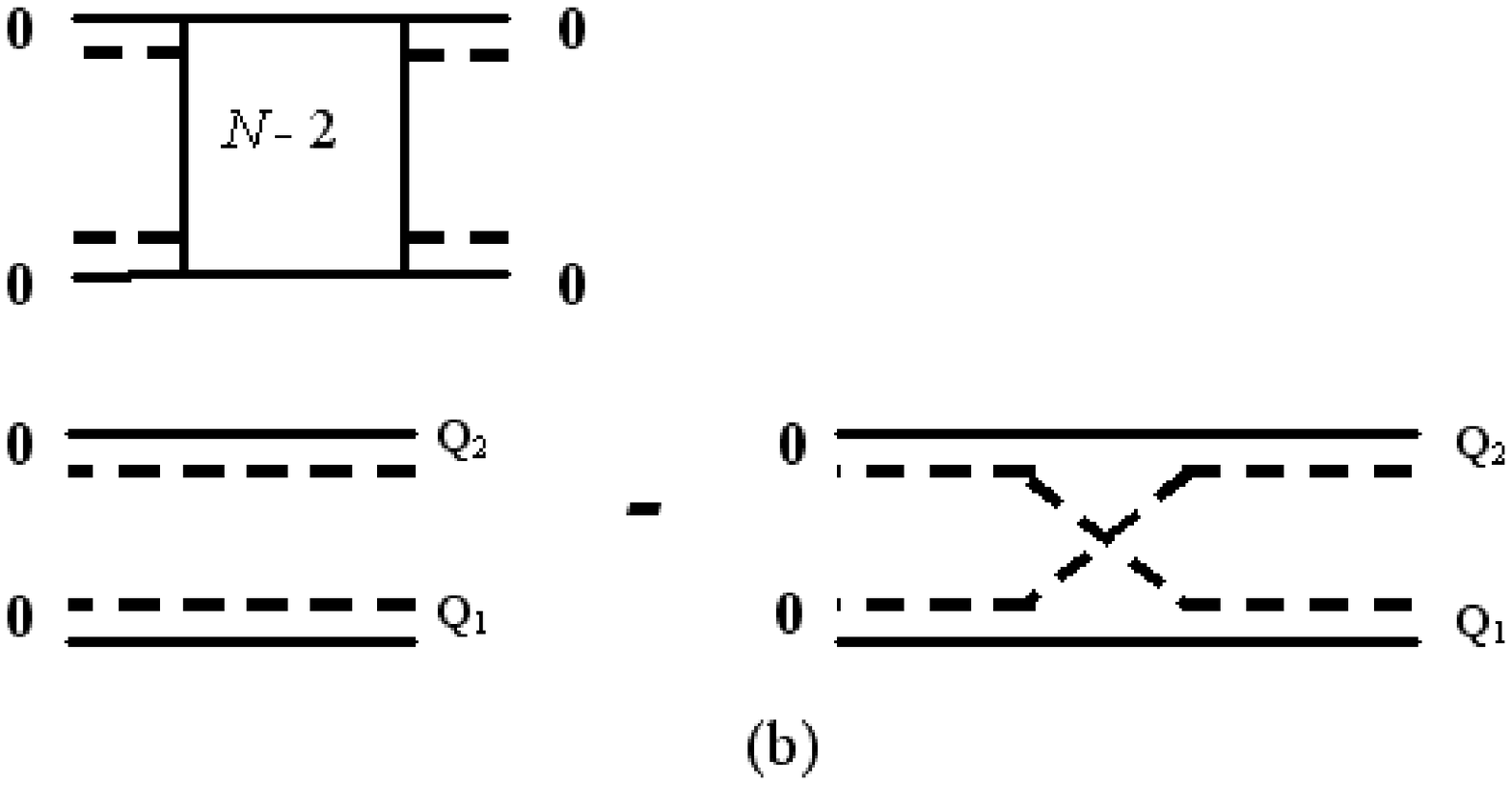}}\\
   \subfigure{\includegraphics[width=0.6\textwidth]{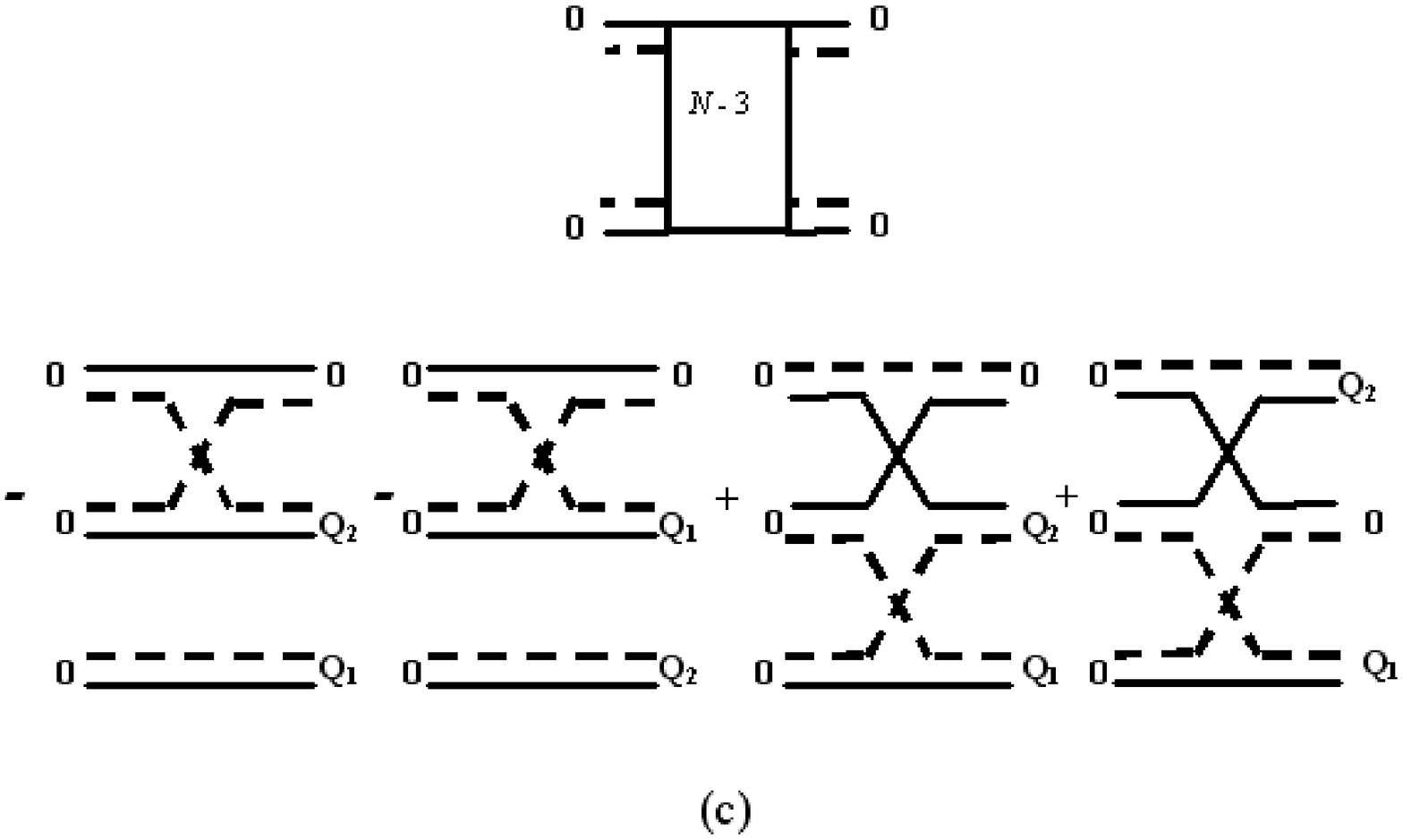}}
\end{center} \caption{\label{Fig5}
(a) Shiva diagrams representation of the scalar product of $N$ exciton state
in which 2 excitons on one side have different momenta $\mathbf{Q}_{1}$ and $%
\mathbf{Q}_{2}$.
(b) First order term: $(N-2)$ excitons $\mathbf{0}$ stay unaffected. The
excitons ($\mathbf{Q}_{1}$, $\mathbf{Q}_{2}$) can either keep their fermion
or have an exchange.
(c) Second order term: $(N-3)$ excitons $\mathbf{0}$ stay unaffected.}
\end{figure}

\end{document}